\documentclass[12pt]{article}
\usepackage{epsfig}
\usepackage[koi8-r]{inputenc}
\usepackage[english,russian]{babel}

\begin{document}

\begin{otherlanguage}{english}
\title{Love story under the H-bomb shadow}
\author{ K.~E.~Filipchuk \\
Novosibirsk State Pedagogical University, \\ 
630 126, Novosibirsk, Russia \vspace*{2mm}\\ 
Z.~K.~Silagadze \\
Budker Institute of Nuclear Physics  and \\ Novosibirsk State
University, 630 090, Novosibirsk, Russia }
\date{}

\maketitle

\begin{abstract}
The physics is created by human beings with all weaknesses 
of human nature. This story caught our attention since it 
demonstrates how fragile the human destiny is and even genius 
cannot find freedom and preserve human dignity in the face 
of the totalitarian state.

\end{abstract}
\begin{center}
\includegraphics[width=70mm]{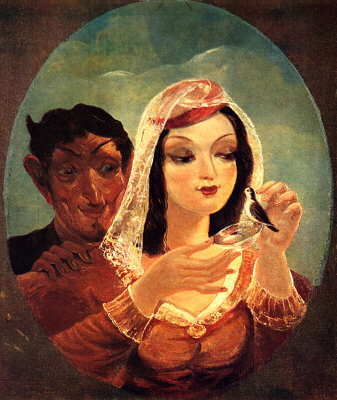}
\end{center}

\section*{}
`` It is not true that in the history of mankind, a great idea 
meant more than a bayonet's spike or howitzer shell. Powerful 
rulers and despots bent into a ram's horn the greatest geniuses 
of their time, and the dark instincts tip the scales much faster 
than the most obvious truth ...'' - so said the character of the 
novel ``Stealing the Moon'' of Georgian writer Konstantine Gamsakhurdia.

The following touching story, which we have learned from the diary of 
a remarkable physicist Mikhail Shifman \cite{1}, adds another detail 
to the broader picture of how the last century's greatest tyrant Stalin 
bent into a ram's horn the whole country. Should he not want to have 
a hydrogen bomb, to bend into the ram's horn the rest of the world too, 
there would be no story. But let's start from a distance.

1969, Moscow, a grand reception in honor of the President of Czechoslovakia, 
Ludwig Svoboda. A nice looking woman, guest at the reception, being a bit 
late, enters deep into the room and appears close to the General Secretary 
Brezhnev. Ludwig Svoboda abruptly drops his talk to Marshal Konev, leaves 
him, runs up to the woman, hugs and kisses her and says ``our daughter?''. 
``No'', - replies the woman, - ``mine.''

The woman's name was Olga. Her and Czechoslovakia's future president's ways 
had crossed in the military 1942, during the formation of the Czech Army. 
At that time, Olga, despite her thirty years, was very silly and naive in her
own words, and she was sure of the infallibility of the Soviet regime. So 
when she was asked to inform the organs about the moods of the Czechs, as she 
did never hide her communications with them, she agreed full of patriotic 
enthusiasm. She dreamt that she would reveal the spy plots. Her husband, who 
worked at the Lubyanka, already did this work before the war. Sometimes she 
and her husband went to a party and then some of those present there were 
arrested. Then, Olga did not attach any importance to this, and only much 
later found out what was it about. 

Olga, of course, did not disclose any spy conspiracy. Maybe because, unlike 
her husband, she could not prevaricate. Instead happened a love-affair with 
Ludwig Svoboda. This was true love. Many years later, she wrote that she thinks
she and Ludwig were halves of each other.

They separated a year later when the Czechs were sent to the front. In 
parting, Ludwig told her "Olia, believe me, I need only you. Soon I'll come 
back for you after we win and we'll be together! My dear, my dear, do not 
forget me. I do not need anybody but you." 

But our story is not about this love affair. Olga had many romantic affairs in 
her life. She was a human being, not an angel, the same being true for heros 
of her affairs as well. Many in that stormy time fell in love and separated 
a lot of times. And after the revolution manners were frivolous. Mother of 
the future secret physicist Janick (so Olga called him), former graduate of 
the Faculty of Philology at the Sorbonne, was terribly shocked by her son's 
manners who drove his first love in his room through the dining room with 
no shame. Janick and Olga would meet after many years at Arzamas-16, in 
a secret atomic center in the Gorky region, where Soviet thermonuclear 
weapons were being developed. But this meeting to take place, the fate still 
had to weave its intricate patterns.

Curators of Olga in the organs were not happy about her performance and 
decided to scare her. Shortly before sending the Czechs to the frontier, Olga 
was approached by her curator Viktor and was asked to follow him. "Instead of 
helping us, you have confused all our cards. I have a warrant for your arrest. 
Look, you intend to go home with bread, but your son is waiting in vain, he 
will not get bread, he will be taken to the orphanage. He is now an 
orphan!" - He said. Olga was not arrested then, but only few years later. She 
was careless in conversation with a mate to call Stalin ambitious and cruel. 
After a while, her friend was arrested, and then Olga. Five years of 
imprisonment for the anti-Soviet agitation. The mate confirmed at the 
confrontation that Olga really called Stalin ambitious and cruel. When Olga 
was interrogated in her turn, six days in a row and all night long without 
sleep, she signed all what they wanted in the end, so she understood 
and forgave her friend. 

Fortunately, in the former life, before the concentration camp, she had time 
to get an education and became a good architect. And in those days prison 
labor was widely used in the Soviet Union for large-scale projects like 
the nuclear one. So she found herself in a secret city of Arzamas-16 and 
began to design a five-story apartment building there. Later, she decorated 
cottages for the bosses and for nuclear physicists. 

Janick's way to Arzamas-16 was different. In 1946, a secret seminar on the 
hydrogen bomb creation problems was held in the U.S. initiated by Teller. 
Soviet government became aware of this seminar at once as the seminar was 
attended by Klaus Fuchs, a Soviet spy who passed atomic secrets to Soviet 
intelligence. Creation of the hydrogen bomb required a lot of calculations 
and theoretical work and for this purpose the Soviet government mobilized 
almost the entire mathematical potential of the Academy of Sciences. The 
young and talented physicist Janick then headed the theoretical division 
of the Institute of Chemical Physics, and he was assigned to coordinate 
the calculations of the hydrogen bomb. So he was sent to Arzamas-16. 
According to the memoirs of Sakharov \cite{2}, Arzamas-16, this strange 
product of that era, ``was a kind of symbiosis of ultramodern Research 
Institute, pilot plants, test sites - and a big concentration camp ... 
Prisoners built plants, testing grounds, roads, housing for future employees. 
They themselves lived in barracks and went to work under an escort of 
accompanied shepherds.'' 

At the beginning, only prisoners with very long terms worked at the object 
that covered a vast area with a solid fence of barbed wire, of which the 
inhabitants of the surrounding poor villages thought that one made ``a trial 
communism'' there \cite{2}. We guess the intention was that they could not 
tell anyone about the object after their release. But one day, about fifty 
of them made a real revolt and were killed in an unequal battle with the 
divisions of the NKVD. After the incident, only prisoners with small terms, 
like Olga, who had something to loose, have been engaged in work at the 
facility. The released prisoners were sent in perpetual exile in Magadan or 
in other similar places where they could not tell anything to anyone. 

So Olga and Janick met in Arzamas-16. The beginning of their romance was 
extravagant. After a date, Janick offered to Olga a ride on a motorcycle. 
But their journey was short; the bike snorted and stood up on a country road. 
Janick looked at his watch, jumped off the bike and ran away, leaving Olga 
among cows. A few days later Janick came back and began to date Olga as 
though nothing had happened. Janick told Olga that he, like Thumbelina 
(he was smaller than Olga) wanted to warm the frozen heart of the 
swallow, so big and beautiful.

And the heart of the swallow melted. On warm summer evenings, they were going 
to the river to look as beavers cut away trees and built dams. They swung on 
a thrown over the river birch drinking vodka with orange juice to keep warm 
when the birch broke and they fell into the water. Olga and Janick appeared 
everywhere together.

Janick wanted a child and Olga became pregnant. It seemed everything was going 
well. After a successful testing of the atomic bomb, Janick received the 
title of hero. Stalin gave him a dacha near Moscow. They got a permission 
to bring Olga's son from her first marriage to her. They had their own house 
where they lived together with Janick. The swallow slowly began to stretch 
her wings.

Once, when Janick went to Moscow as usually, Olga was called by the police 
chief Shutov and he proposed to cooperate, that meant to become an informer. 
Olga refused. ``Think, you will have a child, and yet we can send you there 
where Makar doesn't pasture his calves. What you have experienced in your 
life until now, this is just flowers'' - said Shutov. 

In the evening, going to bed, Olga thought she should accept, for the sake 
of the child. But in the morning, she realized that she could not beat the 
deal with her conscience. She was arrested and two days later sent there 
``where Makar doesn't pasture his calves''. For the two days, during which 
she remained in the area, neither Janick, nor any other of Olga's friends, 
but the son, have visited her. They were afraid. It is true, Janick never 
opposed to the Soviet system, and treated its superhuman strength, most 
likely, with a tremulous delight. At least, that's how Janick remembered, 
according to the memoirs of Sakharov \cite{2}, his dinner with the KGB 
head of the Soviet occupation zone of Germany in 1945, where he was invited 
during his duty journey to Peenem\"{u}nde. The purpose of the business trip 
was to familiarize with the German work on the ballistic missile V-2. 

Soon God's punishment overtook Shutov. Lightning killed his only daughter, 
indeed the beautiful creature, and he himself was dismissed from work. But 
the diabolical carousel in the fate of Olga was already running. In January 
1951, in a crooked corner of the earth, a thousand kilometers from Magadan, 
in the seventy degree frost Olga gave birth to Janick's daughter, the one 
about who Svoboda would ask afterwards whether she was his daughter. There 
was no health post in the village. Only a neighbor, the wife of a local 
bandit Leshka, who once worked in the medical unit, was helping Olga.

Janick was aware of these cases. Though he did not venture to visit Olga 
then in the zone, but he borrowed money from Sakharov and transferred it 
to her. And afterwards he tried to ease her plight. He told Sakharov that 
the floor in the house where Olga gave birth, was covered with a few 
centimeters thick layer of ice. After twenty years, Janick introduced 
to Sakharov his Magadan born daughter at a scientific conference in Kiev. 
Sakharov remembered that Janick was always dreaming to gather his children 
together one day, and he had six of them, from different women. 

Olga was again saved by her specialization. She was transferred into Vytegra 
(a town in the Vologda region, near Lake Onega) in a project team, and 
she joined the welcoming and friendly staff. In 1952, she traveled to Moscow 
and met Janick. Only then she learned that Janick was married, was married 
for a long time, since 1937. Impossibility to continue the relationship with 
a married man was without question. 

Olga lived until 2000, survived Stalin, Janick, and even the Soviet Union, 
the great and terrible country, which suddenly vanished like a smoke on 
a summer day. All of the above we have learned from her memories \cite{3}.

It remains for us to open the cards. If anyone has not guessed yet, the 
secret physicist Janick was Yakov Zeldovich, an exceptionally talented 
physicist, of the level of Landau. The famous British physicist Hawking 
once wrote about him that before he met him, he thought that Zeldovich was 
a nickname of the whole group, and not of a real person, so numerous and 
diverse were his scientific works. 

Yes, he was a genius in physics, but a love story with Olga uncovers his 
human dimension. We cannot say that he was simply a coward. When Sakharov 
was out of favor and was deprived of all his awards, at the international 
congress Zeldovich asked Grischuk to ask him a question after a report 
``Yakov Borisovich, why were you sitting at the plenary session with the stars 
of Hero, but have come here without them?'' he allegedly prepared a joke. 
The question was asked and Yakov Borisovich said ``I did not wear my awards 
for the reason that here is a man who deserves them more than I do, but who 
cannot wear them yet'', expressing in this encrypted form his protest against 
the fact that Sakharov was deprived of his deserved awards \cite{4}.

However, Zeldovich was a very cautious man and did not support Sakha\-rov 
openly, although respected him as a physicist very much. Komberg provides 
a funny story\cite{4}. Once Andrei Dmitrievich delivered a talk at the 
Zeldovich's department about his work on a multi-sheet universe, illustrating 
the story by the models made from sheets of paper glued together. After 
the report Yakov Borisovich asked around for an unnecessary newspaper. 
Everyone thought he would cut his own version of the Multi-sheet world from 
the newspaper. Instead, Zeldovich put the newspaper on the floor near the 
feet of surprised Sakharov, theatrically knelt before him and said: ``Andrei 
Dmitrievich! Well, come on, stop dealing with this nonsense. After all, 
there are very important problems in cosmology that no one but you can solve. 
Well, work at least on quantum gravity.''
 
Yes, the human dimension of great men is not always as perfect as our 
imagination sometimes draws, and maybe some jealous guardians of the other 
people's morality would condemn Janick. But it would be better if we just 
appreciate our current relative freedom, and remember that the ``freedom is 
never more than one generation away from extinction'' (Ronald Reagan, former 
President of the USA). Otherwise, we will be bent again into the ram's horn, 
and will be afraid to visit a pregnant girlfriend, which is sent to the edge 
of the world there ``where Makar doesn't pasture his calves'', for alleged 
misdeeds.
\end{otherlanguage}

\clearpage
\vspace*{5mm}
\centerline{{\LARGE Любовная история на фоне водородной бомбы}}
\vspace*{3mm}
\centerline{{\Large К.~Е.~Филипчук}}
\vspace*{1mm}
\centerline{{\large Новосибирский государственный педагогический университет,}}
\vspace*{1mm}
\centerline{{\large 630 126, Новосибирск}}
\vspace*{2mm}
\centerline{{\Large З.~К.~Силагадзе}}
\vspace*{1mm}
\centerline{{\large Институт Ядерной Физики им. Будкера и}}
\vspace*{1mm}
\centerline{{\large Новосибирский Государственный Университет, 630 090,
Новосибирск}}

\vspace*{5mm}
\begin{abstract}
Физику создают человеческие существа со всеми слабостями человеческой природы. 
Эта история привлекла наше внимание, поскольку она демонстрирует, насколько 
хрупкой является человеческая судьба, и даже гений не может обрести свободу 
и сохранить человеческое достоинство в условиях тоталитарного государства.
\end{abstract}
\begin{center}
\includegraphics[width=65mm]{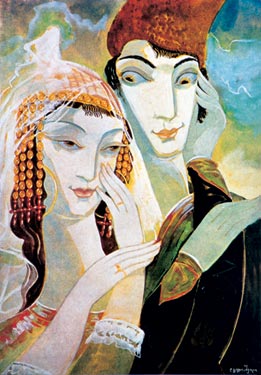}
\end{center}

\section*{}
``Неверно, будто в истории человечества гениальная идея значила больше, нежели 
острие штыка или гаубичный снаряд. Могущественные полководцы и деспоты сгибали 
в бараний рог величайших гениев своего времени; и темные инстинкты решали 
исход дела гораздо быстрее, чем самые очевидные истины...'' - так говорил 
персонаж романа ``Похищение луны'' грузинского писателя Константине 
Гамсахурдия.

Следующая трогательная история, о которой мы узнали из дневника замечательного 
физика Михаила Шифмана \cite{1}, добавляет еще один штрих в картине о том, 
как в 
прошлом веке величайший деспот Сталин согнул в бараний рог целую страну. Не 
захотев он иметь водородную бомбу, чтобы согнуть в бараний рог и весь 
остальной мир, не было бы этой историй. Но начнем издалека.  

1969 год, Москва, большой прием в честь приезда президента ЧССР Людвига 
Свободы. Миловидная женщина, приглашенная на прием, не\-много опоздав, 
проходит вглубь зала и оказывается рядом  с генеральным секретарем Брежневом. 
Людвиг Свобода, беседовавший с маршалом Коневым, бросает его, подбегает 
к женщине, обнимает ее, целует и спрашивает: "Дочь наша?". 
``Нет'', - отвечает женщина, - ``моя''.

Женщину звали Ольга. Их пути с будущим президентом Чехословакии пересеклись 
в военном 1942 году во время формирования чешской армии. В ту пору Ольга, 
несмотря на свои тридцать лет, была очень глупа и наивна, по своим же словам, 
и была уверена в непогрешимости советской власти. Поэтому, когда ей предложили 
информировать органы о настроениях чехов, общения с которыми она не скрывала, 
она  согласилась полная патриотического энтузиазма. Ей грезилось, что она 
будет раскрывать шпионские заговоры. Ее муж, работавший на Лубянке, этим 
уже занимался до войны. Иногда они с мужем ходили на вечеринки, и после этого 
некоторых из присутствовавших там арестовывали. Тогда Ольга не придавала этому 
значения и только позже узнала что к чему. 

Никаких шпионских заговоров Ольга, конечно, не раскрыла. Может, потому, что в 
отличие от мужа она не умела кривить душой. Вместо этого случился у нее роман 
с Людвигом Свободой. Это была настоящая любовь. Спустя много лет она напишет, 
что ей кажется, что они с Людвигом были половинками друг друга. 
 
Они расстались через год, когда чехов отправили на фронт. На прощание Людвиг 
говорил ей: ``Олюша, поверь, ты у меня одна. Скоро я вернусь за тобой с
победой, и мы будем вместе! Милая моя, дорогая, только ты меня не забудь. 
Кроме тебя мне никто не нужен.'' 

Но наш рассказ не об этой любви. Любовных романов у Ольги было много. Она была 
человеком, а не ангелом, впрочем,  как и герой ее романов. Многие в то бурное 
время влюблялись и расставались много раз. И нравы после революции были 
свободные. Маму будущего секретного физика Яника (так будет звать его Ольга), 
выпускницу филологического факультета в Сорбонне, страшно шокировало, что ее 
сын водил первую свою любовь в свою комнату через столовую, никого не 
стесняясь. Яник и Ольга встретятся через много лет в Арзамасе-16, в секретном 
атомном центре в Горьковской области, где разрабатывалось термоядерное оружие.
Но чтобы эта встреча состоялась, судьба еще должна была сплести свои 
хитроумные узоры.

Кураторы Ольги в органах были недовольны и решили ее напугать. Незадолго до 
отправки эшелона чехов на фронт к Ольге подошел ее куратор Виктор Иванович и 
попросил следовать за ним. ``Вместо того, чтобы нам помочь, вы спутали все 
карты. У меня лежит ваш ордер на арест. Вот вы идете с хлебом домой, а ваш сын 
его не дождется, его заберут в детский дом. Он теперь сирота!'' - сказал он. 
Но Ольгу арестовали не тогда, а через несколько лет. Она имела неосторожность 
в разговоре с подругой назвать Сталина властолюбивым и жестоким. Через 
некоторое время подругу арестовали, а потом и Ольгу. Пять лет лагерей за 
антисоветскую агитацию. Подруга на очной ставке подтвердила, что Ольга 
действительно называла Сталина властолюбивым и жестоким. Когда ее саму шесть 
дней подряд допрашивали ночи напролет, не давая спать, в конце концов и Ольга 
подписала все, что от нее требовали; Тогда она поняла и простила подругу.      

К счастью, в прежней, до-лагерной жизни она успела получить образование и
стала хорошим архитектором. А в Советском Союзе тогда широко применялся труд
заключенных на таких масштабных проектах, как атомный. Так она оказалась
в секретном городе  Арзамасе-16 и стала проектировать там пятиэтажный жилой 
дом. Позже она отделывала коттеджи для начальства и физиков-ядерщиков. 

Путь Яника в Арзамас-16 был другой. В 1946 году в США по инициативе Теллера
был организован секретный семинар по проблемам создания водородной бомбы.
Об этом сразу стало известно советскому руководству, так как на семинаре 
присутствовал Клаус Фукс, который передавал атомные секреты советской 
разведке. Создание водородной бомбы требовало много расчетов и теоретической 
работы, и для этой цели советское правительство мобилизовало почти весь 
математический потенциал Академии Наук. Молодой и талантливый физик Яник 
тогда заведовал теоретическим отделом института химической физики, и ему 
поручили координировать работу по расчету водородной бомбы. Так он оказался 
в Арзамасе-16. По воспоминаниям Сахарова \cite{2}, Арзамас-16, это странное 
порождение эпохи, "представлял собой некий симбиоз из сверхсовременного 
научно-исследовательского института, опытных заводов, испытательных 
полигонов - и  большого лагеря...Руками заключенных строились заводы, 
испытательные площадки, дороги, жилые дома для будущих сотрудников. Сами же 
они жили в бараках и ходили на работу под конвоем в сопровождении овчарок."  

В начале на объекте, охватывающим огромную территорию со сплошной оградой из 
колючей проволоки, про который жители окрестных нищих деревень думали, что
там устроили "пробный коммунизм" \cite{2}, работали заключенные с очень 
большими сроками. Наверное, чтобы они после освобождения не могли никому 
рассказать об объекте. Но однажды около пятидесяти из них устроили настоящее 
восстание и погибли в неравном бою с дивизиями НКВД. После этого к работам 
на объекте стали привлекать только заключенных с малыми сроками, наподобие 
Ольги, которым было что терять. Освободившихся ссылали на вечное поселение 
в Магадан или в другие подобные места, где они никому ничего не могли 
рассказать.

Вот так встретились Ольга и Яник в Арзамасе-16. Начало их романа было 
экстравагантным. После знакомства Яник предложил Ольге прокатиться на 
мотоцикле. Но катались они недолго, мотоцикл зафыркал и встал на проселочной 
дороге. Яник посмотрел на часы, спрыгнул с мотоцикла и убежал, оставив Ольгу
среди стада коров. Через несколько дней Яник вернулся и снова стал ухаживать
за Ольгой, как будто ничего не случилось. Яник говорил Ольге, что он, как 
Дюймовочка (он был ниже Ольги ростом), хочет отогреть сердце замерзшей 
ласточки, такой большой и красивой.

И сердце ласточки растаяло. В теплые летные вечера они ходили к речке
смотреть, как бобры спиливали деревья и устраивали плотины. Они качались 
на перекинутой над рекой березе. Пили водку с апельсиновым соком чтобы 
согреться, когда березка подвела и они свалились в воду. Ольга и Яник всюду 
появлялись вместе.

Яник хотел ребенка, и Ольга забеременела. И казалось все шло хорошо. После
успешного испытания атомной бомбы Яник получил звание героя. Сталин ему 
подарил дачу под Москвой. Получили разрешения и привезли сына Ольги от 
первого брака. У них был свои дом, где они жили вместе с Яником. Ласточка 
потихоньку стала расправлять крылья.

Однажды, когда Яник в очередной раз уехал в Москву, Ольгу вызвал начальник 
МВД объекта Шутов и предложил сотрудничать, т.е. стать стукачом. Ольга
отказалась. ``Подумайте, у вас должен быть ребенок, а ведь мы можем вас 
отправить туда, куда Макар телят не гонял. То, что вы пережили, - это 
цветочки,'' - сказал Шутов. 

Вечером, ложась спать, Ольга думала, что должна согласиться ради ребенка. 
Но утром поняла, что не сможет одолеть сделку с совестью. Ее арестовали
и через два дня отправили туда "куда Макар телят не гонял". Те два дня,
в течение которых она оставалась в зоне, ни Яник, ни кто другой из 
Ольгиных знакомых, кроме сына, ее не навестили. Они боялись. Впрочем,
Яник никогда не выступал против советской системы и относился к ее 
нечеловеческой мощи, скорее всего, с трепетным восхищением. По крайнее
мере, именно так вспоминал Яник, по воспоминаниям Сахарова \cite{2}, свои
ужин с начальником ГБ советской зоны оккупации Германии в 1945 году, на
который его пригласили во время его командировки в Пенемюнде. Целью
командировки было ознакомление с немецкими работами по баллистической 
ракете Фау-2. 

Вскоре Божья кара настигает Шутова: его единственную дочь, писаную 
красавицу, убивает молния, а его самого увольняют с работы. Но дьявольская
карусель в судьбе Ольгы уже запущена. В январе 1951 года в Богом забытом
уголке земли, в тысяче километров от Магадана, при  семидесятиградусном 
морозе Ольга рожает дочку Янику, ту про которую потом Свобода будет
спрашивать, не его ли она дочка. В поселке медпункта не было. Ольге 
помогала соседка, жена бандита Лешки, которая когда-то работала в 
медсанчасти.

Яник был в курсе этих дел. Он хоть и от страха не навестил Ольгу тогда
в зоне, но деньги у Сахарова занял и ей передал. А потом старался 
облегчить ее участь. Он рассказывал Сахарову, что в доме, где рожала 
Ольга, пол на несколько сантиметров был покрыт льдом. Через двадцать лет 
Яник познакомил Сахарова со своей родившейся в Магадане дочерью на 
научной конференции в Киеве. Сахаров вспоминает, что Яник мечтал 
когда-нибудь свести вместе своих детей. А их у него было шестеро, от 
разных женщин.

Ольгу опять спасает ее специальность. Ее переводят в Вытегру (город 
в Вологодской области, около Онежского озера) в проектную группу,
и она попадает в приветливый и доброжелательный коллектив. В 1952
году она ездит в Москву и встречает Яника. Только тогда она узнает,
что Яник женат, давно женат, еще с 1937 года. О продолжении отношений 
с женатым мужчиной не могло быть и речи.  

Ольга дожила до 2000 года, пережив и Сталина, и Яника, и даже 
Советский Союз, великую и страшную страну, которая вдруг растворилась,
как дым в летный день. Все вышеприведенное мы почерпнули из  
воспоминаний Ольги \cite{3}.

Осталось открыть карты. Если кто еще не догадался, секретный физик 
Яник - это Яков Борисович Зельдович, талантливейший физик уровня Ландау. 
Известный английский физик Хоукинг как-то про него написал, что до встречи 
с ним он думал, что Зельдович - это собирательное имя целой группы, а не
реальный человек: так многочисленны и разнообразны были его научные
работы.  

Да, он был гением в физике, но любовная история с Ольгой обнажает его 
человеческое измерение. Нельзя сказать, что Зельдович был совсем уж 
трусом. Когда Сахаров был в опале, и его лишили всех наград, на 
международном конгрессе Зельдович попросил Грищука задать ему после 
доклада вопрос ``Яков Борисович, почему Вы на пленарном заседании сидели 
со звездами Героя, а сюда пришли уже без них?'', якобы он подготовил
шутку. Вопрос был задан, и Яков Борисович ответил: ``Я не надел своих 
наград по той причине, что тут присутствует человек, который больше 
меня их достоин, но который носить их пока не может'', выражая в такой
зашифрованной форме свой протест против лишения Сахарова заслуженных им 
наград \cite{4}.

Но Зельдович был очень осторожным человеком и открыто Сахарова не 
поддерживал, хотя очень уважал его как физика. Комберг приводит
забавный случай \cite{4}. Однажды Андрей Дмитриевич рассказывал в отделе 
Зельдовича свою работу о многолистной Вселенной, иллюстрируя свой 
рассказ склеенной из листов бумаги моделями. После доклада  Яков 
Борисович попросил окружающих ненужную газету. Все подумали, что он
сам вырежет из газеты свой вариант многолистного мира. Вместо этого
Зельдович, постелив газету возле ног удивленного Сахарова, театрально 
встал на колени перед ним и произнес: ``Андрей Дмитриевич! Ну, бросьте 
Вы заниматься этой ерундой. Ведь есть очень важные в космологии 
проблемы, которые кроме Вас никто не сможет решить. Ну, займитесь 
хотя бы квантовой гравитацией''.

Да, человеческое измерение великих людей не всегда так безупречно, 
как иногда рисует наше воображение, и, возможно, некоторые ревнивые
радетели за чужую нравственность осудят Яника. Но будет лучше, если
мы просто будем ценить нашу сегодняшнюю относительную свободу и 
будем помнить, что ``свобода никогда не находится более чем в одном 
поколении от вымирания'' (Рональд Рейган, экс-президента США). 
Иначе нас снова согнут в бараний рог, и будем бояться навестить
беременную любимую женщину, которую отправляют на край земли,
``туда, где Макар телят не гонял'' за мнимые прегрешения.

\section*{Acknowledgements}
We have used paintings of Georgian painter Lado Gudiashvili
(1896-1980) at front pages. We are grateful to Olga Chashchina
for her help with the manuscript.

\end{document}